# Distributed Record Linkage in Healthcare Data with Apache Spark

Mohammad Heydari
School of Industrial and Systems Engineering
Tarbiat Modares University
Tehran, Iran
m_heydari@modares.ac.ir

Reza Sarshar
School of Industrial and Systems Engineering
Tarbiat Modares University
Tehran, Iran
reza_sarshar@modares.ac.ir

Mohammad Ali Soltanshahi
School of Industrial and Systems Engineering
Tarbiat Modares University
Tehran, Iran
m.soltanshahi@modares.ac.ir

*Abstract*— Healthcare data is a valuable resource for research, analysis, and decision-making in the medical field. However, healthcare data is often fragmented and distributed across various sources, making it challenging to combine and analyze effectively. Record linkage, also known as data matching, is a crucial step in integrating and cleaning healthcare data to ensure data quality and accuracy. Apache Spark, a powerful open source distributed big data processing framework, provides a robust platform for performing record linkage tasks with the aid of its machine learning library. In this study, we developed a new distributed data matching model based on the Apache Spark Machine Learning library. To ensure the correct functioning of our model, the validation phase has been performed on the training data. The main challenge is data imbalance because a large amount of data is labeled false, and a small number of records are labeled true. By utilizing SVM and Regression algorithms, our results demonstrate that research data was neither over-fitted nor under-fitted, and this shows that our distributed model works well on the data.

## I. Introduction

The healthcare industry generates vast amounts of data daily, including patient records, clinical data, insurance claims, and administrative records. This data is collected by various healthcare providers, insurance companies, and government agencies. Integrating and analyzing these disparate data sources is essential for improving patient care, reducing costs, and conducting medical research. However, data integration in healthcare presents significant challenges due to variations in data formats, missing values, and inconsistent data entry. Record linkage, also known as data matching or deduplication, is a process that identifies and links records that correspond to the same entity across different datasets. It is a critical step in data integration, as it helps eliminate duplicates and ensures data accuracy. In the context of healthcare, record linkage is used to merge patient records from different sources, identify patient readmissions, detect fraud in insurance claims, and create comprehensive patient profiles. Record linkage is a sophisticated and data-intensive technique used to identify and connect related information across multiple datasets. In healthcare, it plays a pivotal role in integrating, cleaning, and harmonizing disparate data sources, with the goal of creating a unified and accurate patient profile. The rationale behind record linkage is clear: patients interact with various aspects of the healthcare system, resulting in fragmented data stored across different systems, departments, and timeframes. As a result, maintaining a comprehensive and up-to-date patient record is vital for improving patient care, enhancing clinical decision-making, and optimizing healthcare operations. Healthcare record linkage involves the intricate process of identifying and associating records that pertain to the same patient, even when those records may contain variations, inconsistencies, or errors. This task is particularly challenging in healthcare due to factors such as typographical errors in patient identifiers, variations in data entry practices, changes in patient demographics, and the need to match records across diverse data sources, including EHRs, insurance claims, billing records, and public health databases. The benefits of effective record linkage in healthcare are multifaceted. Firstly, it enhances patient care by providing a holistic view of a patient's medical history, including diagnoses, medications, treatment plans, and test results. This comprehensive patient profile enables healthcare providers to make more informed decisions, reducing the likelihood of medical errors and improving treatment outcomes. Secondly, it empowers medical researchers with access to robust, linked datasets, facilitating epidemiological studies, clinical trials, and health outcome assessments. Moreover, record linkage plays a pivotal role in healthcare fraud detection and prevention, ensuring that healthcare funds are allocated efficiently and transparently.

## II. Background

Barreto et al. [1] proposed a record linkage approach for Brazilian public health system to produce very accurate data marts for use by statisticians and epidemiologists. They also discussed OpenMP-based implementation for results comparison.

Omar Ali et al. [2] introduces a novel two-step approach for record linkage, emphasizing the generation of high-quality training data. The approach employs the unsupervised random forest model to measure similarity between records, resulting in a similarity score vector for matching. Various constructions are proposed to select non-match pairs for training data, considering balanced and imbalanced distributions. The top and imbalanced construction is identified as the most effective, achieving 100% correct labels. Comparing random forest and support vector machine algorithms, the random forest with the top and imbalanced construction achieves a comparable F1-score to probabilistic record linkage methods. On average, this approach improves the F1-score by 1% and recall by 6.45% compared to existing record linkage techniques. By emphasizing high-quality training data creation, this approach has the potential to enhance the accuracy and efficiency of record linkage in diverse applications.

Aghamohammadi et al. [3] specifically focused on linking two databases, namely the Statistical Center of Iran and the Social Security Organization. Their study highlights the use of record linkage techniques, specifically focusing on the inclusion of variables and features related to Persian texts as part of the database connection process.

Gkoulalas-Divanis et al. [4] presented a comprehensive review of the research literature on privacy-preserving record linkage. It covers the evolution of techniques across different generations, highlighting their strengths and weaknesses. Their study included a taxonomy of methods and an extensive survey of the latest generation of techniques. Furthermore, their study concludes with a roadmap outlining the direction of analytics-driven approaches that aim to tackle the major challenges in the field.

Karakasidis et al. [5] worked on several techniques to enhance the matching capabilities of the Soundex algorithm. These techniques are specifically designed for privacy-preserving record linkage and utilize a protocol based on Apache Spark, making it suitable for processing big data. They evaluate the proposed alternatives in terms of matching quality and time performance. Results demonstrate that their approach achieves precision and recall rates exceeding 95% for large datasets within a matter of seconds, without the need for privacy-preserving blocking techniques.

The objective of Avoundjian [6] study was to assess and compare the performance of record linkage algorithms frequently employed in public health practice. study involved comparing five deterministic record linkage algorithms (exact, Stenger, Ocampo 1, Ocampo 2, and Bosh) and two probabilistic algorithms (fastLink and beta record linkage [BRL]). The simulations revealed that BRL and fastLink consistently achieved high recall rates across different levels of data quality, while demonstrating comparable precision to deterministic algorithms.

Demelius et al. [7] assessed a specialized encoding technique intended for numerical information and adapts it for encoding geographical coordinates within Bloom filters. Synthetic data is employed to compare the suggested numeric encoding approach for geocoordinates with the string-based method they proposed. The suggested technique for encoding geographical coordinates in Bloom filters achieves greater recall and precision compared to the traditional string encoding method.

Christen et al. [8] introduced an innovative encoding method for Privacy-Preserving Record Linkage (PPRL) using autoencoders, which converts Bloom filters (BFs) into real-number vectors. To ensure the comparability of encoded data from various data owners and achieve superior comparison quality of the resulting numerical vectors, they proposed a specific methodology. Experimental results with real-world datasets demonstrate the effectiveness of their technique in achieving high linkage quality while mitigating known cryptanalysis attacks on Bloom filter encoding. Armknecht et al. discussed BF-based PPRL schemes are susceptible to attacks, particularly pattern mining and graph matching attacks. Previous proposals to strengthen these schemes against such attacks either lacked thorough security analysis or compromised efficiency and linkage quality. In contrast, their study proposes an extension to the scheme by incorporating a linear diffusion layer, which effectively addresses these challenges. Extensive theoretical and experimental analysis confirms that the enhanced scheme maintains high efficiency, linkage quality, and significantly improves security against attacks.

Valkering et al. [9] assess the suitability of Apache Spark for scaling Privacy-Preserving Record Linkage (PPRL). They recognized the significance of having a scalable PPRL implementation, and leveraging Spark offers the added benefit of broad deployability and the potential for future ecosystem enhancements. Their findings indicate that a Spark-based PPRL solution surpasses alternative approaches in handling large volumes of records, demonstrates scalability across numerous nodes, and achieves comparable results to conventional record linkage implementations.

Wiegand et al. [10] conducted a comprehensive review of existing literature and commercial solutions to identify examples of similar ideas or methods that address the challenges associated with record linkage problem. They classified the methods for solving the Record Linkage Problem into four distinct categories: a) Deterministic Record Linkage. b) Probabilistic Record Linkage. c) Classification Approaches and d) Collective Matching Approaches.

Baker et al. [11] studied on capability to connect records belonging to the same individual across different datasets key challenge. that difficulty is further compounded by the inclusion of genomic and clinical data in datasets that may extend across multiple legal jurisdictions, as well as the requirement to facilitate re-identification in specific situations. Their study, reports and justifies the findings and recommendations of Privacy-Preserving Record Linkage (PPRL) methods towards the ability to link records associated with a specific individual across diverse datasets.

Franke et al. [12] studied on scaling Privacy-preserving record linkage (PPRL) technique. They introduced parallel PPRL (P3RL) approaches that leverage distributed dataflow frameworks like Apache Flink or Spark. Their proposed approach incorporates blocking techniques, including the utilization of LSH (locality sensitive hashing achieve both high-quality results and scalability. In another related study, Karakasidis et al. [13] proposed a new privacy-preserving blocking technique called Multi-Sampling Transitive Closure for Encrypted Fields (MS-TCEF). This innovative method efficiently filters records based on redundant assignments to blocks, ensuring better fault tolerance and maintaining high-quality results while scaling linearly with the dataset size. They provide a theoretical analysis of the method's complexity and demonstrate its superiority over existing privacy-preserving blocking techniques in terms of recall and processing costs. Marcel et al. [14] assessed a novel parallel Privacy-Preserving Record Linkage (PPRL) technique utilizing Apache Flink, which focuses on achieving high performance and scalability when dealing with large datasets. Their approach incorporates a pivot-based filtering method for metric distance functions, which reduces the number of similarity computations required. They outline distributed methodologies for determining pivots and carrying out pivot-based linkage. They showcase the exceptional efficiency of the approach across various datasets and configurations.

### III. RECORD LINKAGE CHALLENGES

Performing record linkage in healthcare data is a complex task due to several challenges:

A. Data Heterogeneity: Healthcare data comes in various formats, including structured data (e.g., Electronic Health Records), unstructured data (e.g., medical notes and reports), and semi-structured data (e.g., insurance claims). Combining these data types requires specialized techniques.

B. Privacy and Security: Healthcare data often contains sensitive information, such as patient names and medical histories. Privacy regulations like HIPAA (Health Insurance Portability and Accountability Act) require strict safeguards, which can complicate data linkage efforts.

C. Data Quality: Inaccuracies in healthcare data, such as misspellings, abbreviations, and missing values, can lead to incorrect matches or failures to link related records.

D. Scalability: Healthcare datasets can be enormous, making traditional record linkage methods computationally intensive and time-consuming.

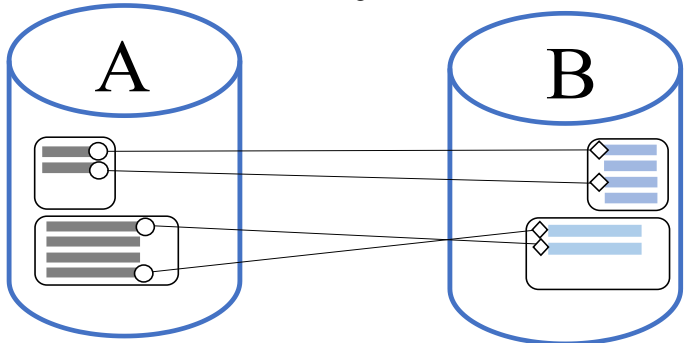

Figure 1 - Record Linkage Process

### IV. APACHE SPARK FOR HEALTHCARE RECORD LINKAGE

Apache Spark is an open-source, distributed data processing framework that excels at processing large-scale datasets efficiently. Spark's machine learning library (MLlib) offers a variety of algorithms for record linkage tasks in healthcare:

A. Data Preparation: Healthcare datasets from different sources are loaded into Spark data frames. Data cleaning and preprocessing steps are performed to standardize fields, handle missing values, and remove noise.

B. Feature Engineering: Relevant features, such as patient names, dates of birth, and addresses, are extracted and transformed into suitable formats for linkage.

C. Record Linkage Model: A machine learning model is trained using MLlib's algorithms to identify matching patient records. This model considers various features and similarity metrics.

D. Evaluation: The linkage model's performance is evaluated using metrics such as precision, recall, and F1-score to ensure accurate record linkage.

### V. RESEARCH METHODOLOGY

Record linkage involves the collection of various datasets from different sources. Once the data processing and standardization phase is complete, the next step is to block and index the data. This allows for efficient comparison of matching records. In the vectorization phase, records are transformed into vectors, and their similarity is classified. Based on this classification, records are placed into three categories: 1) complete matching, 2) lack of matching, and 3) possible matching. Through training and testing processes, we evaluate these categories and calculate key metrics such as precision, recall, and F1-Score. We have defined a roadmap towards achieving these goals, ensuring that the solution can be adjusted and scaled as needed.

A. Dataset

The records in this dataset [15] contain personal information such as first and family names, gender, date of birth, and postal code. These records were collected over several years through iterative insertions. To create comparison patterns, a sample of 100,000 records from the years 2005 to 2008 was used. Each data pair was manually reviewed by multiple documentarists and classified as either a "match" or "non-match". This extensive manual review process served as the basis for evaluating the effectiveness of the registry's own record linkage procedure. To reduce the number of comparison patterns, a blocking procedure was implemented. This procedure selects only record pairs that satisfy specific agreement conditions. The results of six blocking iterations were combined: 1. First name and family name were phonetically equal, and the date of birth was equal. 2. First name was phonetically equal, and the day of birth was equal. 3. First name was phonetically equal, and the month of birth was equal. 4. First name was phonetically equal, and the year of birth was equal. 5. The complete date of birth was equal. 6. Family name was phonetically equal, and the gender was equal. As a result of this procedure, there are 5,749,132 record pairs, out of which 20,931 are considered matches. The dataset is divided into 10 blocks of approximately equal size, ensuring a balanced ratio of matches to non-matches.

Table 1 – Dataset Description

| Feature | Value |
|---|---|
| Characteristics | Multivariate |
| Subject Area | Healthcare |
| Associated Tasks | Classification |
| Attribute Type | Real |
| Instances | 5,749,132 |
| Attributes | 12 |

Here is a description of the attributes in the dataset:

Table 2 - description of the attributes in the dataset

| Attribute | Description |
|---|---|
| id_1 | Internal identifier for the 1st record |
| id_2 | Internal identifier for the 2nd record |
| fn1 | Agreement score for the 1st component of the 1st name |
| fn2 | Agreement score for the 2nd component of the 1st name |
| ln1 | Agreement score for the 1st component of the family name |
| ln2 | Agreement score for the 2nd component of the family name |
| gender | Agreement score for the gender |
| bg | Agreement score for the day component of the date of birth |
| bm | Agreement score for the month component of the birthdate |
| by | Agreement score for the year component of the birthdate |
| plz | Agreement score for the postal code |
| is_match | Matching status (TRUE: match, FALSE: non-match) |

The agreement scores for name components are real numbers ranging from 0 to 1. A score of 0 represents maximum disagreement, while a score of 1 indicates complete equality between the corresponding values. For other comparisons, only the values 0 (not equal) and 1 (equal) are used. The "is_match" attribute is the outcome variable, indicating whether the records are a match or not. The "id_1" and "id_2" attributes are not utilized for prediction purposes but can be used to establish connected components based on the identified matches.

B. Data Loading and Preprocessing

Apache Spark [16] MLlib [17] library is key module for implementing machine learning algorithms. The pyunpack module is used to extract archive files. In this section, multiple datasets are extracted and aggregated from the web address.

C. CSV Datasets Aggregation

By aggregating all datasets into a single dataset, we create a data frame from it and store it on a specific variable. The number of records in the dataset is 5,749,132.

We utilize cache function to store data on the memory.

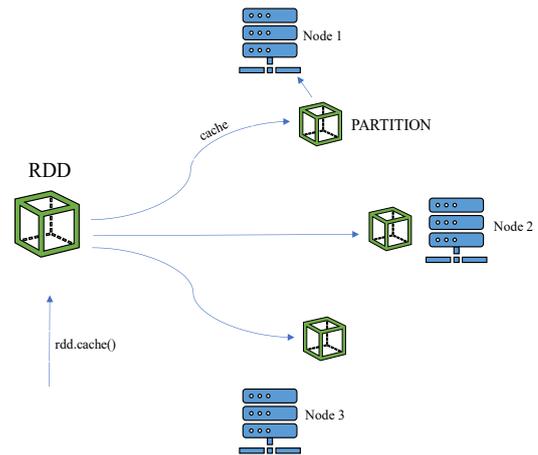

Figure 2 - Cache process in RDD

D. Dataset Records Vectorization

To identify features as feeds for the machine learning algorithm, we use only columns that have less than 20% of the missing values. A threshold is also set where we keep records where at least 3 of its columns have a value and are not empty and deletes the rest. In the *full_count* variable, the total number of records is stored after applying the threshold. training and test data are randomly separated by a ratio of 80 to 20, respectively. We use two functions, *strip* and then *split* on the rows of records. *strip* function Removes all specified characters from the beginning and end of the string. With the *split* function, we can separate sub-strings from the main string based on a specific separator and character.

E. Model Training and Data Sampling

Using the MLlib library, the regression and SVM algorithms are trained on the data, then validation is performed on the training data, and finally the finalized model is executed on the test data. Positive data were isolated by 20931 cases and negative data by 5728201 cases. As it is clear, the imbalance of positive and negative charge in the number of records is quite clear and the need to balance the data is fully felt. We acted on positive data by 0.8 to 0.1 and 0.6 to 0.3 on negative data. In this section, we have used the concept of Fraction Sampling in the range of stratified sampling, which has been the product of the negative record deduction, including 571709 items. In stratified sampling part, random separation of positive training data, positive test data and validated positive data was performed in the range of *0.7, 0.2* and *0.1*, respectively. In the following, random sampling of negative training data, negative test data and validated negative data was performed in a categorized manner with *Fraction = 0.1* and *Seed = 3*.

F. Model Evaluation

To advance the project, the steps to do so have been implemented. In this regard, we have defined the steps towards a dynamic, scalable, and customizable solution.

|       |       |
|-------|-------|
| 4114  | 1109  |
| 3     | 15818 |

Figure 3 - Confusion Matrix of SVM Model

|       |       |
|-------|-------|
| 3942  | 608   |
| 175   | 16319 |

Figure 4 - Confusion Matrix of Regression Model

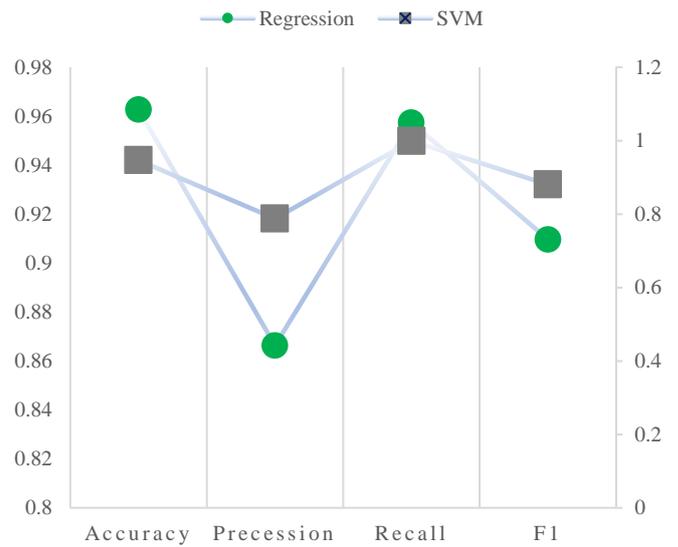

Figure 5 - Algorithms Results Comparison

## VI. RESULTS

Interestingly, both algorithms exhibited certain similarities in their performance metrics. The regression algorithm demonstrated higher accuracy (96.71%) compared to SVM (94.71%). This suggests that the regression model provides a more precise classification of records, which is crucial for healthcare applications. Precision measures the ratio of correctly predicted positive instances to the total predicted positive instances. The regression algorithm achieved a precision of 91.95%, outperforming SVM (78.76%) significantly. This indicates that the regression model is better at minimizing false positives, which is particularly important in healthcare to avoid incorrect patient matches. Recall measures the ratio of correctly predicted positive instances to all actual positive instances. While SVM had a high recall (99.92%), the regression algorithm achieved a recall of 96.65%. Both algorithms demonstrated strong performance in identifying true positive matches. The F1-Score, which balances precision and recall, was higher for the regression algorithm (94.24%) compared to SVM (88.09%). This indicates that the regression model strikes a better balance between minimizing false positives and false negatives.

Table 3 - Algorithms Results Comparison

| *                       | Accuracy | Precession | Recall | F1     |
|-------------------------|----------|------------|--------|--------|
| **SVM**                 | 0.9471   | 0.7876     | **0.9992** | 0.8809 |
| **Regression Validation** | 0.9671 | 0.9195     | 0.9665 | 0.9424 |
| **Regression**          | **0.9627** | **0.8663** | 0.9574 | **0.9096** |

## VII. DISCUSSION

The results of this study highlight the effectiveness of machine learning algorithms with a distributed based approach to data processing in healthcare record linkage. While both SVM and Regression showed promise, the regression algorithm outperformed SVM in terms of accuracy, precision, and F1-score. The higher precision achieved by the regression model is particularly important in healthcare applications, as it reduces the risk of erroneous patient matches. However, the recall achieved by both algorithms indicates their ability to identify a high proportion of true positive matches.

## VIII. CONCLUSION

Record linkage is a crucial process in healthcare data integration, enabling better patient care, research, and cost reduction. Apache Spark, with its machine learning library MLlib, provides an efficient platform for addressing the challenges associated with record linkage in healthcare, such as data heterogeneity, scalability, and privacy concerns. By leveraging Spark's capabilities, healthcare organizations can unlock valuable insights from their data while ensuring data quality and patient privacy. As healthcare data continues to grow, the use of Apache Spark for record linkage will become increasingly essential in optimizing healthcare operations and research. Future research could explore the fine-tuning of hyperparameters, feature engineering techniques, and the application of privacy-preserving methods to enhance the performance of record linkage algorithms in healthcare data. Additionally, the scalability and efficiency of these algorithms in handling larger healthcare datasets should be investigated.